# Emerging Trends for Global DevOps: A New Zealand Perspective

Waqar Hussain, Tony Clear, Stephen MacDonell
*Software Engineering Research Lab (SERL)
School of Engineering, Computer and Mathematical Sciences (SECMS),
Auckland University of Technology (AUT) Auckland, New Zealand*
whussain,tclear,smacdonell@aut.ac.nz

**Abstract**

*The DevOps phenomenon is gaining popularity through its ability to support continuous value delivery and ready accommodation of change. However, given the relative immaturity and general confusion about DevOps, a common view of expectations from a DevOps role is lacking. Through investigation of online job advertisements, combined with interviews, we identified key Knowledge Areas, Skills and Capabilities for a DevOps role and their relative importance in New Zealand's job market. Our analysis also revealed the global dimensions and the emerging nature of the DevOps role in GSE projects. This research adds a small advanced economy (New Zealand) perspective to the literature on DevOps job advertisements and should be of value to employers, job seekers, researchers as well educators and policy makers.*

## 1. INTRODUCTION

DevOps has recently gained popularity as a philosophy that synergizes the operational silos of Software Development (Dev) and IT Operations (Ops) [1]. The three main catalysts that propel its rapid adoption include: a) higher quality expectations from software as it is increasingly offered as a service in the cloud b) demands for rapid delivery of change with growing acceptance of agile and its change embracing attitude, and c) the availability of on-demand powerful and plentiful hardware on the cloud [2]. The uptake of the DevOps trend has been global [1]. Some large organizations claim to have successfully applied its practices in their distributed teams and achieved smooth team collaboration, shortened feedback loops and better customer collaboration [3].

A DevOps strategy supports a globally scalable, rapid and incremental service delivery strategy within a cloud computing infrastructure. Thus it offers potential for software and services companies to operate and compete successfully beyond the traditional centres of technology innovation. For many 'Small Advanced Economies[1] (SAEs) such as New Zealand [4], this service model has attractions. Governments and the IT sector see opportunities to move themselves up the global value chain through delivery of high value products and services. Underpinned by a skilled population base, they believe this will result in more high paying jobs and greater export income [5].

Many believe that DevOps is here to stay, at least in many IT sectors to help organizations deliver quality service with efficiency [2]. However, given the relative recency and emergent nature of the DevOps phenomenon, an inadequate body of knowledge, and general confusion surround DevOps concepts and definitions [6]. The software industry therefore, does not share a common view of its meaning. This lack of clarity fuels several misperceptions. Employers are unable to set and describe the right expectations (Knowledge Skills and Capabilities (KSCs)) for DevOps roles. Job seekers are thus unsure of the commonly assigned responsibilities and expectations [7]. Educators on the other hand, find it challenging to train students and impart the desired skillsets and capabilities for their smooth transition into these roles [8].

A recent study has compared responsibilities of a DevOps engineer role in three countries (USA, UK and Canada) and has shown that the responsibilities and skills expected from DevOps roles significantly vary from country to country [8]. Motivated by country specific differences in their study, this paper analyzes what New Zealand employers actually state they require in DevOps job listings and the extent to which Global Software Engineering (GSE) aspects are included, whether explicitly or implied. The aim is to better understand the growing phenomenon of DevOps in the New Zealand job market, (as one example of an SAE), identify major KSCs and understand how some of those relate to GSE. We draw on interviews and insights from practitioners, to complement the job description data. The questions this research tries to address are:

*1. What knowledge areas, skills and capabilities (KSCs) are in demand for a DevOps role in New Zealand's IT market?*

*2. Which specific KSCs are desired for a DevOps role for distributed development projects?*

---

[1] The Small Advanced Economies include Denmark, Finland, Ireland, Israel, New Zealand, Singapore and Switzerland. These countries have advanced economies that are similar in scale in terms of population (approximately 5 to 10 million). The report cited here has used this 'basket' of economies to benchmark New Zealand performance.

This study will identify implications and chart directions for research and practice in the DevOps area, which may have wider applicability to other small advanced economies. We hope the results will spread awareness regarding the DevOps job landscape in NZ and help job seekers as well as graduates gauge their employability in this area. Our findings might also encourage employers to revisit their approach to better understand trends and express their needs in online job ads. It will also enable educators to better align their current programs with the DevOps job market trends if they choose to do so. For researchers, it opens the venues for further investigations about KSCs, their implications for GSE and the evolving DevOps roles in a global context.

This paper proceeds as follows. Section 2 briefly situates the study in the literature to provide the background and related work in this area. Section 3 describes the research method, which essentially covers the collection and content analysis of job listings, and interview data. Section 4 then presents the results from the study followed by a discussion related to KSCs, DevOps in GSE context and the capabilities desired from DevOps roles working in distributed teams in Section 5. Some issues arising for researchers, educators and practitioners in this area are also discussed in Section 5. Section 6 describes some assumptions and limitations of this work. Finally, Section 7 concludes the paper.

## 2. LITERATURE REVIEW

### A. Background

Traditionally, a disconnect has existed between software development and operations [9], frustrating organizational desires for continuous value delivery. It was apparent that the 'wall of confusion' caused by independent departments working separately could not achieve the levels of productivity and quality demanded by modern day software [10]. A systems thinking approach was therefore needed, one that emphasized organization-wide collaboration to address the lack of cohesion between these departments [11]. The DevOps approach tries to address this problem by continuously integrating software development with operational deployment [12]. It envisions software development work as one activity that flows across functions instead of viewing it as individual actions that remain "lurking in silos" [11]. This is how DevOps as a strategy aims to break down the functional silos and improve collaboration as well as productivity [13].

The DevOps movement gained momentum as more and more professionals began to realize its potential benefits in the world of building, deploying, and maintaining environments [11]. With maturing technology and on-demand plentiful resources in the cloud, the possibilities of a wider industrial application of DevOps was increased. By putting DevOps philosophy into practice, the 'unicorns' of DevOps such as Amazon, Google, Netflix and Snapchat are observed to be achieving significant performance improvements and success [14]. Furthermore, these companies substantially invested into DevOps and have innovatively advanced the release engineering techniques and technologies which can now reduce release cycle times to days or even hours [15]. Several equally successful DevOps model variants such as NoOps, ChatOps, and SmartOps are brought to the fore by these organizations [16, 17].

A recent report [18], quantifies the benefits of investing into the DevOps initiatives. The key findings are: high-performing IT organizations[2] deliver 200 times more frequently, have 2,555 times faster lead time, 24 times faster recovery time with 3 times lower change failure rates, improved quality, security and business outcomes. In another report by Gartner's [19], DevOps is comfortably positioned at the peak of Hype Cycle, leaping from the bottom of the application services trend curve within the last five years (2011-2016). Future predictions are even more promising:

*"By 2016, DevOps will evolve from a niche strategy employed by large cloud providers to a mainstream strategy employed by 25 percent of Global 2000 organizations". [20]*

DevOps is 'trendy' as tempting success stories of DevOps 'Unicorns', SMEs, and startups are echoed from websites, reports, books, blogs, social and broadcast media [14]. Similarly, slogans like *"Enterprise DevOps adoption Isn't Mandatory - but Neither is Survival"* [21] are enticing everyone to jump upon the DevOps bandwagon [1] without fully understanding what it is. As a result, confusion prevails in the industry as to what DevOps actually means or entails when it comes to engaging people with this strategy [6, 8].

Although DevOps emphasizes people (and culture) over tools and processes, actual implementation utilizes technology. Tools are thus considered mandatory to automate DevOps [13]. The common fluid delivery practices of these tools and technologies enable complexity management in software development. DevOps tools however, are growing organically. With literally tens of possible tools to facilitate particular aspects of DevOps practices, practitioners are having to compile visual aids (such as, Automation tool tables [13] a 'Periodic table' of tools [22] to make sense of which tools belong to which family and for what purpose. Currently these tools facilitate fifteen areas spread across software development to deployment (configuration, database, report and release management; Continuous Integration, Deployment, testing, collaboration, cloud, security and so on). Each area or activity has anywhere between six to fourteen tools.

---

[2] "an organization that achieves financial and non-financial results that are exceedingly better than those of its peer group over a period of time of five years or more, by focusing in a disciplined way on that which really matters to the organization."

This situation is well reflected with the booming DevOps toolset market growth that was presaged to reach $2.3 billion in 2015 [Gartner News Room 5 March 2015]. However, the growing complexity in the repertoires of DevOps tools means that there is uncertainty as to what is the best combination of tools. Employers find it difficult to advertise and seek expertise in the right tool combination and job seekers are unsure about the most advantageous and suitable suite of tools to learn for a DevOps role-fit.

**B. Related Work**

Investigation into what employers want from their potential employees could be carried out in a number of ways such as interviews, surveys from employers and job seekers or unobtrusively through online job ads [23]. Several studies have been conducted using these methods in the field of IS using job ads to assess the employer requirements often as an attempt to reconcile the gaps between industry needs and academia's offerings [23, 24].

To the best of our knowledge there are few studies related to DevOps in a GSE context. This has been one of the motivations to carry out this research. Only two studies were found that related to DevOps in the recent International Conference on Global Software Engineering 2016 (ICGSE 2016). One was by Diel et al., [25] describing communication challenges of a distributed agile team due to their geographical, socio-cultural, and temporal distance. The other study by Calefato and Lanubile [26] was on improving distributed team coordination to enable continuous deployment using a *'hub-and-spoke'* model.

While there are several studies using online job postings related to Information Systems (IS), of note are two which closely relate to our research. The first is by Kanij, Merkel and Grundy [24] online advertisements to identify software testing related job responsibilities. They analyzed 47 job postings and found that a tester's responsibilities went beyond test specific activities. They presented a summary of tasks specifically related to testing-specific responsibilities as well as other tasks testers have to perform routinely. They concluded that testers have to perform non test specific additional activities such as research and development, requirements analysis and learning new technologies to satisfactorily complete their job.

Kennan and colleagues [27], carried out a similar study on online job ads to understand market demands for Skills, Knowledge and Competencies (SKC) of early information systems (IS) graduates. They identified patterns in job-specific requirements and they placed under relevant categories. They identified IS Development (a mix of personal competencies and technical skills in software development) as the core combination SKCs desired in early graduates. The second major area of the desired competency in students was more technical in nature requiring skill and competence in Architecture and Infrastructure, Operating Systems, Network and Security.

The most recent study using online job advertisements is by Kerzazi and Adams [8]. They analyzed 211 randomly selected online job postings of five different countries and for three particular roles; DevOps Engineer, Build and Release Engineers. They identified common themes of activities these roles have to perform. They argue that companies do not fully understand what skills to look for when they advertise to hire people in these roles. This, according to them, was due to a lack of common vocabulary and the body of knowledge in the DevOps area; a finding that resonates well with the existing literature especially regarding the opaqueness in the definition of the DevOps concept [28].

There is some similarity in research design between our work and that carried out by Kerzazi and Adams' in [8], therefore it is important that we establish a distinction between the two studies. Several differences bring novelty to our work. First is the difference in focus; Kerzazi and Adams' study is aimed at identifying the common activities performed by DevOps Engineers whereas our research is focused on the understanding the knowledge, technical skills and capabilities required by employers to carry out these activities. Second is the scope of job role, while they were mainly interested DevOps engineer role, our study had a broader scope and included all advertised jobs for any DevOps role. As a result, our job search picked up titles such as Infrastructure & Automation Specialist, Integration Technical Lead, Java Full Stack Web Developer and so on. Furthermore, we included all the 69 jobs identified through our search compared to taking a sample of identified jobs as done by Kerzazi and Adam. Third and the most important difference is the inclusion of GSE element in our study. Unlike in [8] our analysis is specifically focused on the evolving DevOps roles that infuse into onshore and geographically distributed teams.

Fourth, rather than looking at jobs advertised in other parts of the world as done in [8] we focused this research specifically on New Zealand's IT job market. As previously mentioned, the NZ context, as an example, of an SAE, was specifically chosen as KSCs are known to vary from country to country as reported in [8]. Finally, in contrast to the single data source used by Kerzazi and Adam, our online job advertisements data is complemented by practitioner interviews and insights shared by them during their public presentations, through guest lectures to our students.

It is a general perception within the fields of research and industry that DevOps is in the ascendancy. Consistent with the quest for 'continuous value delivery'[7], it is believed to be here to stay, at least in many IT sectors [1, 2]. However, the opaqueness of DevOps, its related practices and the required toolset makes it challenging for employers to describe opportunities and seek candidates with suitable KSCs. It is possible that the academic program development may have evolved somewhat differently to DevOps practice. Therefore, educators remain puzzled with what might be enduring and what should or should not be included in their programs. This mistiness negatively affects their endeavors to adequately train students to enhance their employability for these emerging roles.

Given the current impact and future predictions and the relative confusion as to what DevOps entail, further research is warranted into these aspects of software development in the job market. Furthermore, given the country-specific variations of DevOps responsibilities [8], it is important to understand KSCs in a NZ context, as one example of a small advanced economy, to compare it with study results from other analogous countries.

## 3. RESEARCH METHOD

### A. Research Design

This research employs a qualitative and interpretive design [29] and a mix of inductive and deductive approaches to investigate the phenomenon of interest. For this research, three sources of qualitative data were utilized to understand Knowledge, Skills and Capabilities (KSCs) for DevOps roles in New Zealand. The sources of data include: 1) Online job advertisements, 2) Practitioner Interviews and 3) Practitioner Presentations. We applied Summative Content Analysis (SCA) and Directed Content Analysis (DCA) respectively [30] to inductively analyze qualitative data from job advertisements. Major and minor categories of KSCs were later developed in accordance with the two sources from the literature [31, 32].

For the analysis of interview data, a Thematic Content Analysis technique was applied as described in [33] and data was transcribed using the practical recommendations shared by [34, 35] for systematically handling qualitative data from interviews.

The organization selected to investigate DevOps implementation for this study is a leading NZ stock exchange listed company that provides financial software to small businesses in NZ and globally. For this study a total of 6 interviews were conducted from a variety of roles that included DevOps training manager, Developers, DevOps lead and Tester. These roles were chosen as they were part of the DevOps implementation initiative launched by the selected company around two years ago.

#### 1) Online Job Advertisements Analysis
##### a) Data Source Selection
The first step necessary for data analysis was to identify and select an appropriate source for job postings data. As indicators of the short to mid-term course of market demands, job listings are probably the most readily available sources to understand knowledge, skills and capabilities wanted by employers in a particular field of work [27]. We chose seek.co.nz[3], a dedicated online job posting company in NZ and Australia, as our data source to assess KSCs for DevOps roles.

##### b) Data Extraction
Data was collected between 16 November 2016 and 18 November 2016. A search on the selected job listing site was carried out using *DevOps* as the keyword and *Information & Communication Technology* as the domain classification provided on the website. Since we were trying to search for jobs throughout New Zealand, area specific search criteria such as City, Town or Suburb was not used. Further filtering of retrieved jobs was possible using criteria such as Type of Work *(Full Time, Part Time, Contract/Temp, Casual/Vacation)* but was not applied as it was not deemed necessary or useful. Complete job descriptions of all 69 job postings were downloaded and stored in a single (Master) MS Word document.

The broad scope of search applied to retrieve job advertisements showed the diversity of job titles and suggested an infusion of DevOps roles into various aspects of software development. Preliminary scanning of job titles revealed a variety of roles, ranging from *DevOps Engineer, Cloud DevOps Engineer, Production Support Developer, Lead Integration Technical Lead to Full Stack Developer* among others.

##### c) Data Processing
The master document of job advertisements was then pruned to eliminate unnecessary spaces, links, copied images and duplicate information. From this document 69 individual job documents were created and given a unique number for identification. Upon further screening, 18 jobs were removed as they were found to be unrelated to the DevOps role as they hardly mentioned DevOps or mentioned a *'DevOps experience'* as a plus. Such jobs included *'Software Applications Solutions Architect'*, *'Agile Consultant'*, *'Big Data Engineer'*, *'Business Intelligence Services Team Lead'*, *'Change Manager'*, *'Practice Manager'* and so on. The remaining 51 documents were then imported into Nvivo, a data analysis software package for further analysis [33].

##### d) Data Classification and Analysis
Following a Directed Content Analysis (DCA) approach, first the key words and terms were identified and coded manually. The identified keywords such as *'Duties and Responsibilities'*, *'What Will you be doing?'*, *'The inside word'*, *'What would I be doing?'*, were initially placed under Key Duties and Responsibilities category. Similarly, *'Demonstrated Skill-set & experience'*, *'Essential Skills, Experience in'*, *'Ideally you will be'*, *'Key Technical Stuff in tool-belt'*, *'What do I need?'*, *'What fits?'*, *'What will you have?'*, *'Why you are a legend?'*, were placed under Core Skills. The development of these categories was refined and guided by the prior categorization in [31]. The final categories used to capture key words and concepts were Knowledge Area, Technology and Tools, Programming Languages, Methodology and Capabilities. These steps were followed by determining operational definitions of the developed categories and, in some cases, modifying them based on the work carried out in [31].

Secondly, using Summative Content Analysis (SCA), a quantification of words appearing in the job advertisement was carried out using a data analysis tool, Nvivo [36]. The purpose was to understand the contextual use of an individual word occurrence and ascertain its relative importance in comparison with the other concepts appearing in the advertisements. Words and phrases were thus examined and placed under relevant categories to further develop the KSC picture. Categories developed manually were later verified by word frequency count carried out through Nvivo to avoid any chances of missing important concepts. The analysis approach was generally composed of the main steps of analyze, depict, tabulate and discuss as done in [31].

---

[3] Listed on New Zealand's official website for career search as an online source with "An extensive list of job vacancies from employers across New Zealand and Australia" https://www.careers.govt.nz/job-hunting/job-vacancy- and-recruitment-websites/

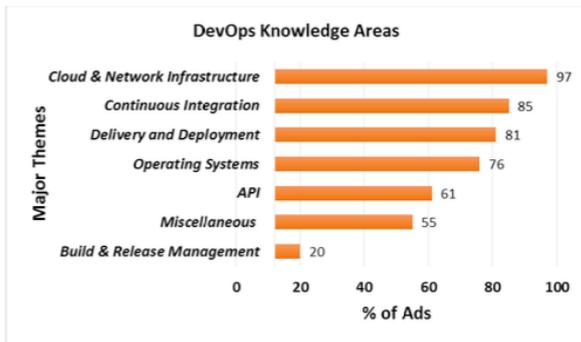

Figure 1: DevOps Knowledge Areas and Major Themes

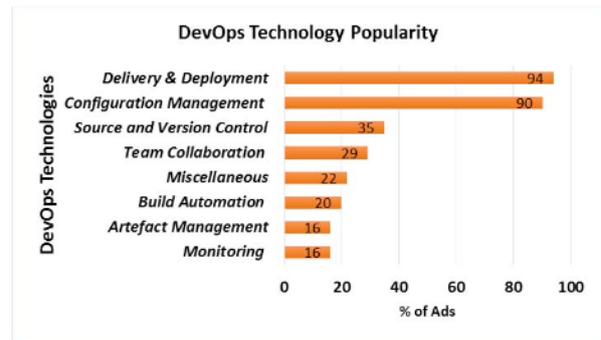

Figure 2: DevOps Technology Popularity

Finally, concepts identified under each main category (along with their occurrence frequency) were grouped into a more representative abstract theme. For example, Build Automation, Build Environment, Build Process were grouped into a *Software Build Process* theme. Similarly, Network Infrastructure (REST, Web services, Firewalls), Network Issues (Trouble Shooting), Network Management (Practice & Procedure, LAN/WAN) were unified under a more abstract and boarder concept Networking. On the basis of this analysis a table of categories, themes and concepts was developed (See Appendix 1).

### 2) Interview Data Analysis

Interviews were transcribed using a Denaturalized approach [35] to retain the accuracy of the interview substance (meaning and perceptions created or shared during the interview).

Transcriptions were re-checked to verify their alignment with the principles of the transcription protocol presented in [34]. Analysis of the interviews was carried out using the Thematic Content Analysis (TCA) technique. Themes were identified in the transcripts using the initially developed KSC categories in this study as well as the general theme of study i.e., 'challenges and benefits of DevOps'.

## 4. RESULTS

Our analysis of online job advertisements revealed major themes and their relative importance under the categories comprising knowledge, skills, and capabilities. We discuss these results in the following sections individually and complement the findings with insights from our interviews and practitioners' presentations.

### A. Knowledge Areas

The most important Knowledge Area (KA) with the highest percentage (97%) is the Cloud & Network Infrastructure. DevOps initiatives are known to utilize cloud technologies and cloud based infrastructure services, so it was not surprising to see knowledge about Amazon Web Services (AWS) and AWS CloudFormation as highly desired KAs across all jobs. The next closely contested knowledge areas are Continuous Integration and Delivery and Deployment with a significantly high percentage of 85 and 81 percent respectively. These figures reflect a general view in the literature that DevOps initiatives are almost inseparable from the drive to gain speed in production and delivery through cloud technologies & CI and CD practices (See Figure 1).

A high percentage of job ads required the candidates to have in-depth knowledge about Unix / Linux operating systems and their administration. A reason for this could be that most of the jobs advertised were for a DevOps Engineer role with an implicit desire for an Ops person with strong knowledge about Operating systems and infrastructure. Among the significant themes APIs Web Services Design and Development were seen as required knowledge areas from the candidates. It may be due to the belief that APIs are an efficient means to programmatically integrate Cloud resources in DevOps[37]. Surprisingly build and release management activities came up only in twenty-one percent of the Ads. One reason could have been that these activities are considered to be the responsibility of developers [38]. Another noteworthy insignificant activity was Testing and Test Automation in the figure which is covered in the Miscellaneous category. The concepts of Cloud, CI and CD are so closely knitted together with test automation in the literature for example [1] that it was surprising to see a lack of focus on Testing Automation (8%) in the advertisements analyzed in this study.

### B. DevOps Technologies

In DevOps technologies, experience with Delivery and Deployment (D& D) tools were the most sought after skill in potential candidates. Hands on experience with tools such as Docker, Jenkins, Bamboo and Octopus were highly desirable as shown from the very high percentage (94 %) for D&D category in Figure 2. Experience with Configuration Management tools were the second most sought after skill job seekers. Tools such as Puppet, Chef, Ansible, and BitBucket were repeatedly mentioned. This means that tools like Puppet and Chef that support CM have maintained their popularity over the last few years in the literature and industry as reported in [39] and [40]. Source and version control tools such as Git and Subversion were apparently less significant compared to Deployment and CM tools with only 35 percent of ads reporting them.

As seen in Figure 2, technological enablers of DevOps such as Build Automation, Artefact management and Monitoring did not appear frequently scoring 20, 16 and 16 percent respectively. Again, this lack of emphasis on Build Automation and Artefact management tools such as (Ant, Gradle, Maven, Artifactory etc.), could be due to the belief that these tools are mainly for supporting developers rather

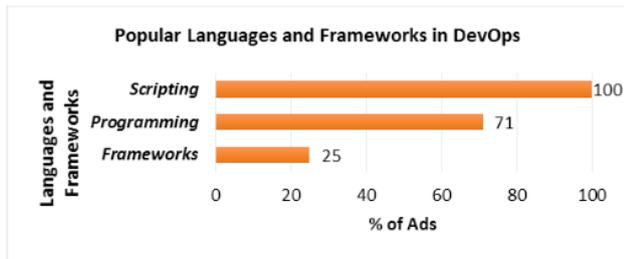

Figure 3: DevOps Languages and Frameworks

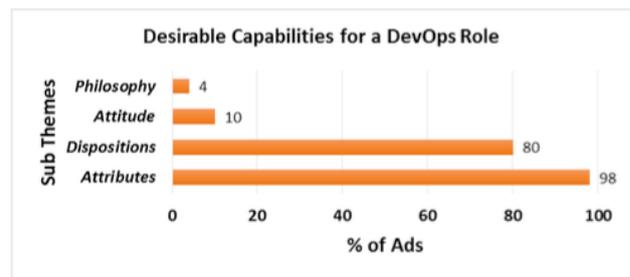

Figure 4: Capabilities Desired for a DevOps Role

than Ops people and hence not required in many of the DevOps engineering jobs.

**C. Languages and Frameworks**
In programming languages, experience with Java, C# and Ruby were commonly desired with 71 percent of the ads requiring competency in these languages (Figure 3). Some form of scripting skills were demanded from the candidates in all advertisements (100%).

The analysis revealed that Python, PHP, PowerShell, JavaScript, Bash and Angular were in high demand (See Appendix 1 for details). Our results confirm scripting related findings by Kerzazi and Adam [8], who also identified scripting tasks to be the most important theme across DevOps roles they evaluated. The commonly cited frameworks of choice for the employers included .NET and Spring however they were specified only in 25 percent of the ads (See Figure 3).

**D. Capabilities**
Complementing the required knowledge and skills, job ads also showed a desire for recruits who had specific (Attributes, Dispositions, Attitude & Philosophy) detailed in Appendix 1. Schussler observes that

> "dispositions are different from knowledge and skills" and "concerns not what abilities people have, but how people are disposed to use those abilities" [41].

The dispositions and attributes sought were typically human and team centric, demanding flexibility and adaptability, a wider mindset including customer/business awareness, relationship management, communication, general business acumen and respect in collaborative relationships (Figure 4).

**E. Job Titles and Roles**
From our analysis of Job listings, we see a diversity of job titles were advertised to join organizations in various DevOps roles (See Appendix 1). Furthermore, job titles whether emphasizing a Dev or Ops aspect, demanded skills extending from Dev to Ops and vice versa. For example, 20% of the job ads sought Full Stack Developers in C#, Java PHP and .NET but also required experience in CI / CD, Build & Release, Cloud technologies and infrastructure management in Linux or Unix.

Similarly, 70% of the job emphasizing Ops elements required the candidates to have strong scripting skills and knowledge and experience of DevOps tools as well as awareness of CI and CD practices. This is an indicator of the industry trend where the DevOps philosophy has started to take its roots where the departmental silos are being challenged. Dev and Ops people are increasingly joining teams and sharing responsibilities from development to deployment of code.

## 5. DISCUSSION

The results from our analysis of job advertisement data raised several interesting points for further discussion from which we can draw a set of conclusions. We discuss our main findings below.

**A. State of DevOps and GSE**
From our analysis we see DevOps occupying an intermediate role in today's NZ software companies. We are seeing differing constellations of teams situated within wider groupings or "tribes"[42], with a typical formation having multiple teams interfacing to a single DevOps release team, often with a Release Manager role who acts as a boundary spanner between the development activities and the production release. So while we are seeing continuous integration we are not seeing continuous deployment yet.

However, supported by our interview data we believe DevOps is evolving from the current restricted position of quite 'bottlenecked tribes' to a more open model of contribution which we might term 'free-wheeling tribes'. In that model teams are distributed but share development and operations roles and all teams can equally continuously integrate and deploy. The role of the release manager in such scenarios becomes defunct.

**B. Absence of Global**
From our data we saw that some 16 percent of the job ads explicitly mentioned global aspects. So we found there is not a strong focus on GSE. However, we also drew the conclusion that global aspects or global dimensions were implicit in most of the advertisements, whether by the cloud context enabling software to be readily deployed globally, the ability and need to serve customers in global markets or need to manage or work collaboratively work with distributed teams.

The capabilities desired by the employers as reported in Appendix give an indication of DevOps roles' orientation towards a distributed or even global nature. The attributes such as Mentorship (14), Customer Engagement (14), Collaboration (8), Leadership (16), Interpersonal Skills (8), Knowledge Sharing (4) Relationship Building (2), Team Player (6), Respectful (1), as noted in Appendix 1, advertised by many companies having global clientele does allude towards a possible distributed or global DevOps role.

### C. Knowledge Areas

Knowledge areas demanded were a mix of traditional development and cloud computing knowledge and skills, based within an infrastructure and technology context. Build, release CI & CD automation practices were sought as critical skillsets. Operating systems and system administration know-how were important, as were APIs and Packaged or high level products supporting web interoperability and enabling productivity, and component and services composition.

During the interviews the DevOps team lead identified a general lack of skillset in networking and infrastructure concepts in people seeking DevOps roles in their organization.

> "We have very hard time hiring for people because the skillset does not exist, like people with good fundamentals of networking and data centre engineering."

According to the DevOps lead, fundamentals of networking, typology diagrams and the understanding of network infrastructure and micro services were essential for any role related to DevOps.

In summary the knowledge areas demanded an extended development skillset comprising both application and infrastructure, configuration, maintenance, performance tuning and scheduling. In addition, an extended skillset for operations personnel comprised, system administration, networking and scripting to maintain, enhance and manipulate virtual infrastructure environments.

### D. Tools and Technologies

In addition to the broader issues associated with DevOps and GSE we identified some major thematic areas which represent the context and needs of employers for DevOps personnel.

In the technology area we saw very rich and complex constellations of technologies being demanded. The picture presented was one of a sophisticated repertoire of tools to be orchestrated in order to perform DevOps functions. This richness and complexity brought its own significant learning overhead.

One of the Developers reported that getting to grips with complexity of DevOps technologies in combination with DevOps practices during the software development lifecycle was quite challenging. The team lead of this developer confirmed the challenges and added that:

> "Her learning curve was like 90 degrees"

We saw specialists, typically senior technical leads, joining teams for periods of time to select or recommend, install, configure and establish working practices around specific tools sets. In a sense we could term the experts in these roles as 'Tribal Nomads'.

Similar insights were shared by industry experts during presentations at our university in NZ talking about how DevOps philosophy has started to take root in the local IT industry. They suggested that often DevOps leads move in and out of the cross functional teams as a bridging role to train and upskill people so that team members share both Dev and Ops responsibilities. As a result, the roles such as operations engineering and release engineers may well be disappearing.

Findings from our study allude to the global dimensions of a DevOps role which is sometimes expressed explicitly but more often it is implied. A possible reason could be that companies in NZ are increasingly using outsourcing as a business model and working in a globally distributed setting and distributed DevOps tribes. The interviewees in the studied organization also discussed DevOps team formations being characterized by free moving members into distributed teams that form 'guilds' ("an organic and wide-reaching "community of interest", a group of people that want to share knowledge, tools, code, and practices" [42]). Based on our study insights we believe that such 'freewheeling tribes' are expected to become more apparent in the near future. This alludes to the GSE element and a trend towards a global DevOps role.

### E. Languages and Frameworks

Again with the languages and frameworks, we saw a diverse set of technologies being employed. In the development roles we saw traditional development languages (3GL's etc.), sometimes within wider vendor frameworks. Web oriented languages were common and from a more operational focus we saw the dominance of scripting languages enabling DevOps personnel to perform system administration and infrastructure management/tuning roles/tasks.

For employers, programming language skills such as those were required in combination with the underlying computing and networking knowledge. In the interviews, the training manager mentioned the absence of this combination in the fresh graduates seeking a DevOps role:

> "When I talk to grads to find out what people are into ... lot of people like Java and front end stuff but do they understand underlying computing system? That's like No!"

### F. Capabilities

A broader theme that came through from analyzing the data related to the attributes and dispositions demanded of DevOps roles, was that leadership attributes were actively sought. The diagram below illustrates Quinn's leadership model, identifying four quadrants with broad profiles of activity. From our data we saw the strong emphasis on 'Transformational Leadership' roles such as Mentor, Facilitator, Innovator and Broker. These stood in contrast to the more traditionally viewed expectations of technical employees engaged in projects and task related activities, better fitting a 'Transactional Leadership' style. Expectations of taking responsibility for other team members through training and mentoring them to develop new KSCs were very evident in the job ads.

Interview data complemented the findings from job advertisement analysis and confirmed the need for broader set of capabilities and dispositions. Analysis of the interviews revealed that in addition to technical skills, for DevOps roles certain personal capabilities and dispositions were also highly valued in the industry such as analytical

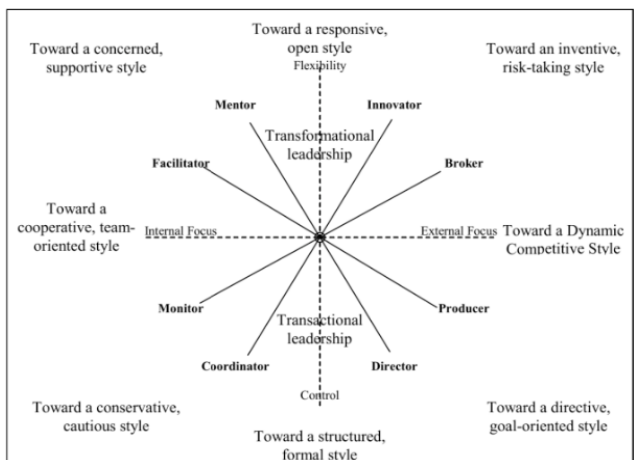

Figure 5: Competing Values Framework of Leadership Roles [43]

ability, creativity, learnability and flexibility to adapt. One of the leads stressed that

> "You have to have quite creative mind-set to be able to build out these [virtual environments] ... it is in that weird spatial and cognitive space that a lot of people may struggle with... Understanding why things are the way they are and how things have evolved is really the key to understanding where things are going and how the industry is moving forward".

**G. GSE Related Job Tasks**

From the responsibilities that explicitly mentioned GSE related tasks, we saw DevOps framed in a context that expected skills in collaboration applying agile approaches, both across functions and across teams in distributed settings aiming towards continuous global deployment. Some explicit examples of a global DevOps role's responsibilities advertised in the job listings included *"working in conjunction with the global continuous delivery team"* and *"working on exciting global projects work across different teams"* as well as *"collaborate and interact with teams in different locations"*.

In addition to these several other capabilities that alluded to a GSE role were highly desired. For example, a candidate's ability to collaborate with multiple teams, lead and mentor team members and engage stakeholders through effective communication and interpersonal skills as noted in Appendix 1. Several job postings also noted a global clientele and/or globally distributed offices and required the candidates to be able to effectively collaborate further hinting towards a global DevOps role.

Based on the most frequently occurring topics within each theme in Appendix one a sample job advertisement could be sketched out for a (global) DevOps role in the NZ IT industry. This might be used by employers, job seekers and educators to align their efforts in seeking, training and educating for major KSCs required in a NZ or possibly in a similar SAE context.

## 6. ASSUMPTIONS AND LIMITATIONS

The main assumption of this research is that content of online job postings is a valid representation of the needs of employers from job seekers. We acknowledge that some job descriptions may not be entirely clear or accurate on the account that they are listed by recruitment agencies or employers who may not be experts in that area or have clarity around what is required. We limited our study to DevOps job advertisements from only one listing company, however we did not specify any special roles within this area nor undertook any samples of the jobs listed. We analyzed all jobs listed under Information and Communication Technology search through the keyword DevOps between 16 November 2016 and 18 November 2016.

This research has utilized a content analysis technique and an unobtrusive method of analyzing online job listing with the belief that a fairly substantial assessment of the market demands could be achieved with quite limited resources. We, therefore, do not claim complete coverage of DevOps jobs in New Zealand as related jobs listed at other venues (word of mouth, career webpages employers) might have been missed. However, we believe that job listings from a widely used and publicly available online employment facilitation platform is likely to uncover a fairly representative list of candidate KSCs in the market.

## 7. CONCLUSIONS

This investigation has provided a snapshot of Knowledge areas, Skills and Capabilities (KSCs) to gain a reasonably comprehensive overview of employers' expectations from DevOps roles in a NZ context. Findings from this study reveal that DevOps is being adopted as a philosophy whereby responsibilities of individual members are shared across increasingly joint Development and Operations teams. As a result, extended capabilities are desired from DevOps roles that go beyond their usual Dev or Ops strengths. This we believe will make more traditional roles such as dedicated Release Engineers fade away.

Furthermore, a GSE perspective is clearly visible as the advertised jobs actively seek dispositions of individuals to collaborate and communicate with distributed stakeholders as well as to lead and mentor team members. The study indicates that the global dimensions of a DevOps role are apparent in almost all of the advertisements sometimes by explicit mention but more often by implication. This seems to represent the state of the art in NZ, where increasingly firms are working in a globally distributed setting and constellations of distributed DevOps teams in 'freewheeling tribes' are expected to become more apparent in near future. This alludes to the GSE element and a trend towards a global DevOps role.

We expect our findings to aid employers and job seekers to provide a general view of the key KSCs to advertise and prepare for respectively. We also expect our findings to guide curriculum developments through options such as including DevOps aspects within existing courses, and creating new courses or even programs (perhaps at the postgraduate level). There are however, some questions that remain unanswered and warrant further investigation:

Are we expecting too much from employees in these roles spanning from Ops to Sys Admin to Developers? How typical is the NZ experience of DevOps industry compared to other countries in Small Advanced Economies and how different is it from large unicorn DevOps settings? What are the drivers and patterns for DevOps take-up which are unique to NZ IT compared to other countries?

In summary, we believe that DevOps is a key trend, underpinned by business drivers of continuous value delivery, and globally scalable technology platforms. However, the demands of the roles required, the inherent organizational tensions, and the ideal global configurations of such "tribes" of DevOps personnel are still emerging. In this study we have investigated DevOps from a New Zealand perspective. Yet, DevOps can definitely be seen as a global phenomenon in software engineering, the implications of which are still becoming apparent.

APPENDIX 1: KNOWLEDGE AREAS, SKILLS AND CAPABILITIES FOR (GLOBAL) DEVOPS ROLES

| CATEGORIES/THEMES | TOPICS/SUB THEMES | | |
|---|---|---|---|
| **Knowledge Areas** | **Topics** | **Freq** | **% Ads** |
| Cloud & Network Infrastructure | AWS & AWS CloudFormation (26), Cloud Technologies (14), Network Infrastructure (9), Azure (7), IaaS & PaaS (6), Middleware Stack (3), Cloud Infrastructure (2), Cloud Environment (1) | 68 | 97% |
| Continuous Integration | CI (53), Pipeline Management (2), Enablement (1) | 56 | 85% |
| Delivery & Deployment | Continuous Delivery (25), Continuous Deployment (24), | 87 | 81% |
| Operating Systems | Linux Administration and Management (27), Unix Administration and Management (17), Windows and MS Active Directory (6) | 50 | 76% |
| API | REST API (11), Web API (6), Design & Development (4), Platform Service (4), Platform Support (4), Micro services (2), SOAP (3), JSON (2), Reporting (1), | 37 | 61% |
| Miscellaneous | Software Development (6), Technical Support (5), Testing (4), Configuration Management (4), Middleware Stack (3), Database (2), ICT Services (2), Change Management (1), Data Store (1), E Commerce (1), Incident Management (1), Performance Matrix (1), Team Experience (1) | 32 | 55% |
| Release & Build | Release Management (5), Release Environment (1), Build Management (6) | 6 | 20% |
| **Technology** | **Tools** | **Freq** | **%Ads** |
| Configuration Management | Puppet (21), Chef (12), Ansible (7), BitBucket (6), Salt (1), OpsWorks (1) | 48 | 90% |
| Delivery, Deployment and Release | Docker (12), Jenkins (11), Bamboo (10), Octopus (7), Team Services (5), Kubernetes (3), Codeship (1) | 49 | 94% |
| Source and Version Control | Git (9), Other Source Control (9) | 18 | 35% |
| Team Collaboration | Atlassian Suite (6), JIRA (3), Confluence (3), Visual Studio (2), Slack (1) | 15 | 29% |
| Build Automation | Ant (3), Gradle (3), Team City (3), Vagrant (1) | 10 | 20% |
| Monitoring | Logstash (3), Nagios (3), Elastic (ELK) 2 | 8 | 16% |
| Artefact Management | Maven (4), Artifactory (3), MongoDB (1) | 8 | 16% |
| Miscellaneous | Cucumber (3), HP UFT (1), Nginx (1), Oracle VM (1), Vmware (1), Apigee (1), MuleSoft (1), Wordpress (1), Joomla (1) | 11 | 22% |
| **Languages & Frameworks** | **Topics** | **Freq** | **%Ads** |
| Programming | Java (17), C# (10), Ruby (10), XML (2), Gherkin (1) | 40 | 71% |
| Scripting | Python (14), PHP (11), Power Shell (9), Bash (8), JavaScript (8), Angular (6), Perl (6), Node.JS (5), CSS (4), mysql (4), HTML5- (3), Jasmine (1), Junit (1), PHPUnit (1), postgresql (1), jQuery (1), BootStrap (1) | 83 | 100% |
| Frameworks | .NET (14), Spring (3), Entity (1) | 18 | 25% |
| **Capabilities** | **Sub Themes** | **Freq** | **%Ads** |
| Attributes | Communication Skills (39), Leadership (16), Customer Engagement (14), Mentorship (14), Adaptability and Learnability (11), Collaboration (8), Interpersonal Skills (8), Problem Solving (7), Analytical Ability (6), Self-Management (6), Handle Pressure (5), Planning (3), Task Management (2), Independence (2), Relationship Building (2), Knowledge Sharing (4) | 147 | 98% |
| Dispositions | Passionate (8), Team Player (6), Motivated (3), Enthusiastic (3), Visionary (3), Curious (3), Innovative (3), Proactive (2), Energetic (2), Confident (2), Technical Orientation (2), Business Acumen (2), Committed (2), Self-Starter (1), Responsible (1), Talented (1), Respectful (1), Pragmatic (1), Favor Free (1), Flexible (1), Hardworking (1), Change Tolerant (1), Quality Conscious (1), Resilient (1), Critical Thinker (1) | 53 | 80% |
| Attitude | can do attitude (2), try anything once (2), won't mind getting hands dirty (1), | 5 | 10% |
| Philosophy | Healthy disregard for functional silos (2) | 2 | 4% |
| **GSE Related Tasks** | **Responsibilities Explicitly Mentioned** | **Freq** | **%Ads** |
| Examples | "working on exciting global projects work across different teams" (1), "collaborate and interact with teams in different locations" (1), "get stuff done across multiple business units" (1), "providing support across the rest of the teams whilst linking up with our internal development teams" (1), "working collaboratively with project & operation teams, embedded within developer teams" (1), "Within true Agile cross-functional teams" (1), "managing Agile software and / or Dev Ops work teams and processes" (1), "working in conjunction with the global continuous delivery team" (1) | 8 | 16% |
| **Roles** | **Job Titles** | **Freq** | **%Ads** |
| Examples | (Senior) DevOps Engineer (20), (Full Stack) Developers (C#, .NET, JAVA, PHP) (10), Build Deploy/Release Developer/ Engineer (3), Software Support/ Operations/ Performance Engineer (3), Unix/ Linux (Support) Engineer - AWS - DevOps (3) DevOps Specialist/Capability Lead (2), Solutions Consultant (DevOps context) (2), Systems/ Support Engineer (2), Others (6) [Development & Operations Batman with Utility Belt, Infrastructure & Automation specialist, Integration Technical Lead, Network Engineer, Production Support Developer, Software Development Coordinator - (Senior Role)] | 51 | 100% |